# ON THE COMPUTATIONAL CAPABILITIES OF PHYSICAL SYSTEMS
# PART I: THE IMPOSSIBILITY OF INFALLIBLE COMPUTATION

by David H. Wolpert

NASA Ames Research Center, N269-1, Moffett Field, CA 94035, dhw@ptolemy.arc.nasa.gov

**Abstract**: In this first of two papers, strong limits on the accuracy of physical computation are established. First it is proven that there cannot be a physical computer C to which one can pose any and all computational tasks concerning the physical universe. Next it is proven that no physical computer C can correctly carry out any computational task in the subset of such tasks that can be posed to C. This result holds whether the computational tasks concern a system that is physically isolated from C, or instead concern a system that is coupled to C. As a particular example, this result means that there cannot be a physical computer that can, for any physical system external to that computer, take the specification of that external system's state as input and then correctly predict its future state before that future state actually occurs; one cannot build a physical computer that can be assured of correctly "processing information faster than the universe does". The results also mean that there cannot exist an infallible, general-purpose observation apparatus, and that there cannot be an infallible, general-purpose control apparatus. These results do not rely on systems that are infinite, and/or non-classical, and/or obey chaotic dynamics. They also hold even if one uses an infinitely fast, infinitely dense computer, with computational powers greater than that of a Turing Machine. This generality is a direct consequence of the fact that a novel definition of computation  — a definition of "physical computation" —  is needed to address the issues considered in these papers. While this definition does not fit into the traditional Chomsky hierarchy, the mathematical structure and impossibility results associated with it have parallels in the mathematics of the Chomsky hierarchy. The second in this pair of papers presents a preliminary exploration of some of this mathematical structure, including in particular that of prediction complexity, which is a "physical computation analogue" of algorithmic information complexity. It is proven in that second paper that either the Hamiltonian of our universe proscribes a certain type of computation, or prediction complexity is unique (unlike algorithmic information complexity), in that there is one and only version of it that can be applicable throughout our universe.

**INTRODUCTION**

Recently there has been heightened interest in the relationship between physics and computation ([1-33]). This interest extends far beyond the topic of quantum computation. On the one hand, physics has been used to investigate the limits on computation imposed by operating computers in the real physical universe. Conversely, there has been speculation concerning the limits imposed on the physical universe (or at least imposed on our models of the physical universe) by the need for the universe to process information, as computers do.

To investigate this second issue one would like to know what fundamental distinctions, if any, there are between the physical universe and a physical computer. To address this issue this first of a pair of papers begins by establishing that the universe cannot contain a computer to which one can pose any arbitrary computational task. Accordingly, this paper goes on to consider computer-indexed subsets of computational tasks, where all the members of any such subset *can* be posed to the associated computer. It then proves that one cannot build a computer that can "process information faster than the universe". More precisely, it is shown that one cannot build a computer that can, for any physical system, correctly predict any aspect of that system's future state before that future state actually occurs. This is true even if the prediction problem is restricted to be from the set of computational tasks that can be posed to the computer.

This asymmetry in computational speeds constitutes a fundamental distinction between the universe and the set of all physical computers. Its existence casts an interesting light on the ideas of Fredkin, Landauer and others concerning whether the universe "is" a computer, whether there are "information-processing restrictions" on the laws of physics, etc. [10, 19]. In a certain sense, the universe is more powerful than any information-processing system constructed within it could be. This result can alternatively be viewed as a restriction on the universe as a whole — the universe cannot support the existence within it of a computer that can process information as fast as it can.

To establish this result concerning prediction of the future this paper considers a model of computation which is actually general enough to address the performance of other computational tasks as well as prediction of the future. In particular, this result does not rely on temporal orderings of events, and therefore also establishes that no computer can infallibly predict the *past* (i.e., perform retrodiction). So any memory system must be fallible; the second law cannot be used to ensure perfectly faultless memory of the past. Accordingly, the psychological arrow of time is not inviolate [24].[1] The results are also general enough to allow arbitrary coupling of the computer and the external universe. So for example they also establish that there cannot be a device that can take the specification of any characteristic of the external universe as input and then correctly observe the value of that characteristic. Similarly, they establish that here cannot be a device that can take the specification of any desired value of a characteristic of the external universe as input and then induce that value in that characteristic. In other words, loosely speaking, there cannot be either an infallible general purpose control device nor an infallible general purpose control device. (This second interpretation can be viewed as an uncertainty principle that does not involve quantum mechanics.)

There are a number of previous results in the literature related to these results of this paper. Many authors have shown how to construct Turing Machines out of physical systems (see for example [10, 22] and references therein). By the usual uncomputability results, there are properties of such systems that cannot be calculated on a physical Turing machine within a fixed allotment of time (assuming each step in the calculation takes a fixed non-infinitesimal time). In addition, there have been a number of results explicitly showing how to construct physical systems whose future state is non-computable, without going through the intermediate step of establishing computational universality [13, 23].

There are several important respects in which the results of this paper extend this previous work. All of these previous results rely on infinities of some sort in physically unrealizable systems (e.g., in [23] an infinite number of steps are needed to construct the physical system whose future state is not computable). In addition, they all assume one's computing device is no more powerful than a Turing machine. Also none of them are motivated by scenarios where the compu-tation is supposed to be a prediction of the future; in fact, they are not in any sense motivated by consideration of the temporal relation between one's information and what is to be predicted. Nor are they extendable to allow arbitrary coupling between the computer and the external universe, as (for example) in the processes of observation and control.

There are other limitations that apply to many of these previous results individually, while not applying to each and every one of them. For example, in [23] it is crucial that we are computing an infinite precision real number rather than a "finite precision" quantity like an integer. As another example, many of these previous results explicitly require chaotic dynamics (e.g., [7]).

None of the limitations delineated above apply to the result of this paper. In particular, no physically unrealizable systems, chaotic dynamics, or non-classical dynamics are exploited in this paper. The results also hold even if one restricts attention to systems which are finite, i.e., which contain a finite number of degrees of freedom. In all this, the results of this paper constitute a novel kind of physical unpredictability.

The results also hold even if the computer is infinitely dense and/or infinitely fast (in which case the speed of light would be infinite). The results also hold even if the computer has an infinite amount of time to do the calculation. They also hold even if the computer's initial input explicitly contains the correct value of the variable it is trying to predict, and more generally they hold regardless of the program running on the computer. They also hold for both analog and digital computation, and whether or not the computer's program can be loaded into its own input (i.e., regardless of the computational universality of the computer). Moreover, the results hold regardless of the power of one's computer, so long as it is physically realizable. If it turns out to be physically possible to have infinitely fast, infinitely dense computers, with computational power greater than that of a Turing machine, then the result of this paper holds for such a computer. As a particular example, the results also hold even if the "computer" includes one or more human beings. So even if Penrose's musing on quantum gravity and intelligence turns out to have some validity, it is still true that human intelligence is *guaranteed* to be wrong sometimes.

The way that this paper derives results of such generality is to examine the underlying issue from the broad perspective of the computational character of physical systems in general, rather than that of some single precisely specified physical system. The associated mathematics does not directly involve dynamical systems like Turing machines. Rather it casts computation in terms of partitions of the space of possible worldlines of the universe. For example, to specify what input a particular physical computer has at a particular time is to specify a particular subset of all possible worldlines of the universe; different inputs to the computation correspond to different such subsets. Similar partitions specify outputs of a physical computer. Results concerning the (im)possibility of certain kinds of physical computation are derived by considering the relationship between these kinds of partitions. In its being defined in terms of such partitions, "physical computation" involves a structure that need not even be instantiated in some particular physically localized apparatus; the formal definition of a physical computer is general enough to also include more subtle non-localized dynamical processes unfolding across the entire universe.

Section 1 of this paper generalizes from particular instances of real-world physical computers that try to "reliably and ahead of time predict the future state of any system" to motivate a broad formal definition of such computers in terms of partitions. To maintain maximum breadth of the analysis, we do not want to restrict attention to physical computers that are (or are not) capable of self-reference. As an alternative, as elaborated at the end of Section 1, we restrict attention to universes containing at least two such physical computers. (Put another way, our results hold for any single computer not so powerful as to preclude the possible existence anywhere else in the universe of another computer as powerful as it is — which certainly describes any computer that human beings can ever create.) Section 2 begins by proving that there exist prediction problems that cannot even be posed to one of those two physical computers. Restrictions on the set of prediction problems are introduced accordingly. Section 2 then proves that, even within such a restricted set of prediction problems, one cannot have a pair of computers each of which can, reliably and ahead of time, predict the future state of any system. This is even true if one allows the possibility that the computer is initialized by having the correct prediction provided to it.

In their most abstract formulation, the results of this section concern arbitrary calculations a computer might make concerning the physical universe, not just "predictions" per se. This is how they also establish (for example) the necessarily fallible nature of retrodiction, of observation, and of control. These results are all derived through what is essentially a physical version of a Cretan Liar's paradox; they can be viewed as a physical analogue of Godel's Incompleteness Theorem, involving two instances of the putative computer rather than self-referential computers.

These issues are addressed in the second of these two papers which, while building on the results of the first paper, is meant to be self-contained. The second papers begins with a cursory review of these partition-based definitions and results of the first paper. Despite its being distinct from the mathematics of the Chomsky hierarchy, as elaborated in that paper, the mathematics and impossibility results governing these partitions bears many parallels with that of the Chomsky hierarchy. Section 2 of that second paper explicates some of that mathematical structure, involving topics ranging from error correction to the (lack of) transitivity of computational predictability across multiple distinct computers. In particular, results are presented concerning physical computation analogues of the mathematics of Turing machines, e.g., "universal" physical computers, and Halting theorems for physical computers. In addition, an analogue of algorithmic information complexity, "prediction complexity", is elaborated. A task-independent bound is derived on how much the prediction complexity of a computational task can differ for two different reference universal physical computers used to solve that task. This bound is similar to the "encoding" bound governing how much the algorithmic information complexity of a Turing machine calculation can differ for two reference universal Turing machines. It is then proven that one of two cases must hold. One is that the Hamiltonian of our universe proscribes a certain type of computation. The other possibility is that, unlike conventional algorithmic information complexity, its physical computation analogue is unique, in that there is one and only version of it that can be applicable throughout our universe.

Throughout these papers, $\mathbf{B} \equiv \{0, 1\}$, $\Re$ is defined to be the set of all real numbers, '^' is the logical *and* operator, and 'NOT' is the logical *not* operator applied to $\mathbf{B}$. To avoid proliferation of symbols, often set-delineating curly brackets will be used surrounding a single symbol, in which case that symbol is to taken to be a variable with the indicated set being the set of all values of that variable. So for example "$\{y\}$" refers to the set of all values of the variable y. In addition $o(A)$ is the cardinality of any set A, and $2^A$ is the power set of A. $u \in U$ are the possible states of the universe, and $\hat{U}$ is the space of allowed trajectories through U. So $\hat{u} \in \hat{U}$ is a single-valued map from $t \in \Re$ to $u \in U$, with $u_t \equiv \hat{u}_t$ the state of the universe at time t. Note that since the universe is microscopically deterministic, $u_t$ for any t uniquely specifies $\hat{u}$. Sometimes there will be implicit constraints on $\hat{U}$. For example, we will assume in discussing any particular computer that the space $\hat{U}$ is restricted to worldlines $\hat{u}$ that contain that computer. An earlier analysis addressing some of the issues considered in this pair of papers can be found in [30].

## I. A DEFINITION OF WHAT IT MEANS TO "PREDICT THE FUTURE"

### i) Definition of a Physical Computer

For the purposes of this paper, a physical computer will "predict the state of a system ahead of time" if the computer is a general emulator of the physical dynamics of such a system, an emulator that operates faster than that dynamics. So given some time $T > 0$, and given some desired information concerning the state of some system at T, our goal is to have the computer output that desired information *before time T*. To that end we allow the computer to be "initialized" at time 0, with different "input", depending on the value of T, what information is desired, perhaps information about the current state of the state whose future is being predicted, etc.

To make this concrete, we start by distinguishing the specification of what we want the computer to calculate from the results of that calculation. Let $\alpha$ be the value of a variable delineating some information concerning the state of the physical universe at time T (e.g., the values of a finite set of bits concerning the state of a particular system S residing in the universe at that time). We indicate a specification that we wish to know $\alpha$ as a *question* $q \in Q$. So q says what $\alpha$ is for any state of the universe at time T. This means that what we wish our computer to tell us is the result of q, a single-valued mapping from the state of the universe at T to α.

Since $\hat{u}$ fixes $u_T$ and (for a deterministic universe) vice-versa, we can generalize this by dispensing with specification of T. In other words, we can recast any q as any single-valued mapping from $\hat{u}$ to α. So q fixes a partition over the space $\hat{U}$, and any pair (α, q) delineates a region in $\hat{U}$.

In general, the space {α} of "potential answers" of the universe (i.e., the set of partition element labels) can change depending on q, the "question concerning the universe" (i.e., the partition). We will write the space {α} as A(q) when we need to indicate that dependence explicitly. We formalize all this as follows:

**Definition 1:** Any *question* q ∈ Q is a pair, consisting of a set A of *answers* and a single-valued function from $\hat{u} \in \hat{U}$ to α ∈ A. A(q) indicates the A-component of the pair q.
Here we restrict attention to Q that are non-empty and such that there exist at least two elements in A(q) for at least one q ∈ Q. We make no other *a priori* assumptions concerning the spaces {A(q ∈ Q)} and Q. In particular, we make no assumptions concerning their finiteness.

Without the accompanying q, a value of α, by itself, is meaningless. So we must know what q was when we read the computer's output. Accordingly, we take the output of our computer to be a question q together with an associated prediction for α. (If the question is only stored in a human user's memory, then that aspect of the human is implicitly part of the computer.) So our computer's output is a delineation of a subregion of $\hat{u} \in \hat{U}$; those $\hat{u}$ such that q( $\hat{u}$ ) = α. Choose some real number τ, where 0 < τ < T. Our goal is that for any q ∈ Q there is an associated initial "input" state of the computer which ensures that at time τ our computer's output is a correct prediction for α, in that for the $\hat{u}$ of the universe, q( $\hat{u}$ ) = α.

**Example 1 (conventional prediction of the future):** Say that our universe contains a system S external to our computer that is closed in the time interval [0, T], and let u be the values of the elements of a set of canonical variables describing the universe. α is the t = T values of the compo-nents of u that concern S, measured on some finite grid G of finite precision. q is this definition of $\alpha$ with G and the like fully specified. (So q is a partition of the space of possible $u_T$, and therefore of $\hat{U}$, and $\alpha$ is an element of that partition.) Q is a set of such q's, differing in G, whose associated answers our computer can (we hope) predict correctly.

The input to the computer is implicitly reflected in its t = 0 physical state, as our interpretation of that state. In this example (though not necessarily in general), that input specifies what question we want answered, i.e., which q and associated T we are interested in. It also delineates one of several regions $R \subseteq \hat{U}$, each of which, intuitively, gives the t = 0 state of S. Throughout each such R, the system S is closed from the rest of the universe during t $\in$ [0, T]. The precise R delineated further specifies a set of possible values of $u_0$ (and therefore of the Hamiltonian describing S), for example by being an element of a (perhaps irregular) finite precision grid over $U_0$, G'. If, for some R, q($\hat{u}$) has the same value for all $\hat{u} \in R$, then this input R uniquely specifies what $\alpha$ is for any associated $\hat{u}$. If this is not the case, then the R input to the computer does not suffice to answer question q. So for any q and region R both of which can be specified in the computer's input, R must be a subset of a region $q^{-1}(\alpha)$ for some $\alpha$.

Implicit in this definition is some means for correctly getting the information R into the computer's input. In practice, this is often done by having had the computer coupled to S sometime before time 0. As an alternative, rather than specify R in the input, we could have the input contain a "pointer" telling the computer where to look to get the information R. (The analysis of these papers holds no matter how the computer gains access to R.) In addition, in practice the input, giving R, q, and T, is an element of a partition over an "input section" of our computer. In such a case, the input is itself an element of a finite precision grid over $\hat{U}$, G". So an element of G" specifies an element of G (namely q) and element of G' (namely R.)

Given its input, the computer (tries to) form its prediction for $\alpha$ by first running the laws of physics on a $u_0$ having the specified value as measured on G', according to the specified Hamiltonian, up to the specified time T. The computer then applies q(.) to the result. Finally, it writes this prediction for $\alpha$ onto its output and halts. (More precisely, using some fourth finite precision grid

G''' over its output section, it "writes out" (what in interpret as) its prediction for what region in U the universe will be in at T, that prediction being formally equivalent to a prediction of a region in $\hat{U}$.) The goal is to have it do this, with the correct value of $\alpha$, by time $\tau < T$.

Consider again the case where there is in fact a correct prediction, i.e., where R is indeed a subset of the region $q^{-1}(\alpha)$ for some $\alpha$. For this case, formally speaking, "all the computer has to do" in making its prediction is recognize which such region in the partition q that is input to the computer contains the region R that is also input to the computer. Then it must output the label of that region in q. In practice though, q and R are usually "encoded" differently, and the computer must "translate" between those encodings to recognize which region $q^{-1}(\alpha)$ contains R; this translation constitutes the "computation".

Consider a conventional computer that consists of a fixed physical dynamical system together with a pair of mappings by which some of that system's observable degrees of freedom are interpreted as (perhaps binary) "inputs", and some as "outputs". More precisely, certain characteristics of the degrees of freedom of the computer — like whether they exceed a pre-specified threshold, in the case of a digital computer — are interpreted that way. The input and output degrees of freedom can overlap, and may even be identical. Since the computer exists in the physical universe its state is specified by u. Therefore both the interpretation of some of the computers degrees of freedom as "inputs" and some as "outputs" is equivalent to a mapping from $u \in U$ to a space of inputs and of outputs, respectively. All of this holds whether the computation of the outputs from the inputs proceeds in a "digital" or "analog" fashion.

Under the convention that the initialization of the computer occurs at t = 0, since $\hat{u}$ fixes $u_0$ and vice-versa, we can broaden the definition of a computer's input, to be a mapping from $\hat{u} \in \hat{U}$ to a space of inputs. So for example "initialization" of a computer as conventionally conceived, which sets the t = 0 state of a physical system underlying the computer, is simply a special case. In that special case, the value taken by the input mapping can differ for $\hat{u}$ and $\hat{u}'$ only if the t = 0 state of the computer portion of the universe, as specified by $\hat{u}$, differs from the t = 0 state of the com-puter portion of the universe as specified by $\hat{u}'$. Similarly, under the convention that $\tau$ is fixed, we can broaden the definition of a computer's output to be a mapping from $\hat{u} \in \hat{U}$ to a space of outputs.

We define a computer by formalizing these considerations:

**Definition 2: i)** A (computation) *partition* is a set of disjoint subsets of $\hat{u}$ whose union equals $\hat{U}$, or equivalently a single-valued mapping from $\hat{U}$ into a non-empty space of partition-element labels. Unless stated otherwise, any partition is assumed to contain at least two elements.

**ii)** In an *output partition*, the space of partition element labels is a space of possible "outputs", {OUT}.

**iii)** In the current context, where we are interested in prediction, we require {OUT} to be the space of all pairs {$OUT_q \in Q$, $OUT_\alpha \in A(OUT_q)$}, for some Q and A(.) as defined in Def. (1). This space — and therefore the associated output partition — is implicitly a function of Q. To make this explicit, often, rather than an output partition, we will consider the full associated double (Q, OUT(.)), where OUT(.) is the output partition $\hat{u} \in \hat{U} \to OUT \in$ {$OUT_q \in Q$, $OUT_\alpha \in A(OUT_q)$}. Also, we will find it useful to use an output partition to define an associated ("**p**rediction") partition, $OUT_p(.) : \hat{u} \to (A(OUT_q(\hat{u}), OUT_\alpha(\hat{u}))$.

**iv)** In an *input partition*, the space of partition element labels is a space of possible "inputs", {IN}.

**v)** A (*physical*) *computer* consists of an input partition and an output partition double. Unless explicitly stated otherwise, both of those partitions are required to be (separately) surjective.

Since we are restricting attention to non-empty Q, {OUT} is non-empty. We say that $OUT_q$ is the "question posed to the computer", and $OUT_\alpha$ is "the computer's answer". The surjectivity of IN(.) and OUT(.) is a restriction on {IN} and {OUT}, respectively. It reflects the fact that, for reasons of convenience, we don't allow a value to "officially" be in the space of the computer's potential inputs (outputs) if there is no state of the computer that corresponds to that input (out-put). For example, if the computer is a digital workstation with a kilobyte of its RAM set aside as input, it makes no sense to have the input space contain more than $(2^8)^{1024}$ values.

While motivated in large measure by the task of predicting the future, the definition of physical computation is far broader, concerning any computation that can be cast in terms of inputs, questions about physical states of nature, and associated answers. This set of questions includes in particular any calculation that can be instantiated in a physical system in our universe, whether that question is a "prediction" or not. All such physically realizable calculations are subject to the results presented below.

Even in the context of prediction though, the definition of a physical computer presented here is much broader than computers that work by the process outlined in Ex. 1 (and therefore the associated theorems are correspondingly further-ranging in their implications). For example, the computer in Ex. 1 has the laws of physics explicitly built into its "program". But our definition allows other kinds of "programs" as well. Our definition also allows other kinds of information input to the computer besides q and a region R (which together with T constitute the inputs in that example above). We will only need to require that there be *some* t = 0 state of the computer that, by accident or by design, induces the correct prediction at t = $\tau$. This means we do not even require that the computer's initial state IN "accurately describes" the t = 0 external universe in any meaningful sense. Our generalization of Ex. 1 preserves analogues of the grids G (in Q(.)), G" (in IN(.)) and G"' (in OUT(.)), but not of the grid G'.

In fact, our formal definition of a physical computer broadens what we mean by the "input to the computer", IN, even further. While the motivation for our definition, exemplified in Ex. 1, has the partition IN(.) "fix the initial state of the computer's inputs section", that need not be the case. IN(.) can reflect *any* attributes of $\hat{u}$. An "input" — an element of a partition of $\hat{U}$ — need not even involve the t = 0 state of the physical computer. In other words, as we use the terms here, the computer's "input" need not be specified in some t = 0 state of a physical device. Indeed, our definition does not even explicitly delineate the particular physical system within the universe that we identify with the computer. (A physical computer is simply an input partition together with an output partition.) This means we can even choose to have the entire universe "be the computer". For our purposes, we do not need tighter restrictions in our definition of a physical computer. Nonetheless, a pedagogically useful example is any localized physical device in the real world meeting our limited restrictions. No matter how that device works, it is subject to the impossibility results described below.

We can also modify the example presented above in other ways not involving input. For example, we can have $T < 0$, so that the "prediction" is of the past. We can also have S be open (or perhaps even be the entire universe), etc. Although prediction of the future is an important (and pedagogically useful) special case, our results hold more generally for any calculation a physical computer might undertake.

As a final example of the freedom allowed by our definition, consider again conventional computation, where both the input and output mappings reflect the state of the portion of the universe consisting of some physical system underlying the computer. Now in practice we may want to physically couple such a computer to the external universe, for example via an observation apparatus that initializes the computer's inputs so that they reflect information about the system being predicted. Such a coupling would be reflected in $\hat{u}$. If we wish though, we can exploit the freedom in its definition to modify the input mapping, in such a way that it too directly reflects this kind of coupling. For example, under the proposed modification, if we want the input section of the computer's underlying physical system to be a bit $b_1$ that equals the $t = -1$ state of some bit $b_2$ concerning the external universe, then we could have IN($\hat{u}$) = IN($b_1(u_0)$, $b_2(u_{-1})$) = $b_1(u_0)$ if $b_1(u_0) = b_2(u_{-1})$, and have it equal a special "input error" value otherwise. If we do have a physical coupling mechanism, and if that mechanism is reliable — something reflected in $\hat{u}$ — then this third setting will never occur, and we can ignore it. However use of this modified IN allows us to avoid explicitly identifying such a mechanism and simply presume its existence. So long as the third setting never occurs, we can analyze the system *as though* it had such a (reliable) physical coupling mechanism.

We will sometimes find it useful to consider a "copy" of a particular computer C. Intuitively,

this is any computer C' where the logical implications relating values of IN' and OUT' are the same as those relating values IN and OUT, so that both computers have the same input-output mapping.

**Definition 2 (v):** Given a computer $C \equiv \{Q, IN, OUT\}$, define the *implication* in $\{OUT\}$ of any value $IN \in \{IN\}$ to be the set of all $OUT \in \{OUT\}$ consistent with IN, in that $\exists\ \hat{u} \in \hat{U}$ for which both $IN(\hat{u}) = IN$ and $OUT(\hat{u}) = OUT$. Then the computer $C^2 \equiv \{Q^2, IN^2, OUT^2\}$ is a *copy* of the computer $C^1 \equiv \{Q^1, IN^1, OUT^1\}$ iff $Q^2 = Q^1$, $\{IN^2\} = \{IN^1\} \equiv \{IN\}$, $\{OUT^2\} = \{OUT^1\}$, and the implication in $\{OUT^2\}$ of any $IN \in \{IN\}$ is the same as the implication in $\{OUT^1\}$ of that IN.

Note that we don't require that $IN^1(.) = IN^2(.)$ in the definition of a copy of a computer; the two computers are allowed to have different input values for the same $\hat{u}$. Conversely, any computer is a copy of itself, a scenario in which $IN^1(.)$ does equal $IN^2(.)$.

## ii) Intelligible computation

Consider a "conventional" physical computer, consisting of an underlying physical system whose $t = 0$ state sets $IN(\hat{u})$ and whose state at time $\tau$ sets $OUT(\hat{u})$, as in our example above. We wish to analyze whether the physical system underlying that computer can calculate the future sufficiently quickly. In doing so, we do not want to allow any of the "computational load" of the calculation to be "hidden" in the mappings $IN(.)$ and $OUT(.)$ by which we interpret the underlying physical system's state, thereby lessening the computational load on that underlying physical system. Stated differently, we wish both the input and the output corresponding to any state of the underlying physical system to be "immediately and readily intelligible", rather than requiring non-trivial subsequent computing before it can be interpreted. As will be seen in our formalization of this requirement, it is equivalent to stipulating that our computer be flexible enough that there are no restrictions on the possible questions one can pose to it.

One way to formalize this intelligibility constraint would entail imposing capabilities for self-reference onto our computer. This has the major disadvantage of restricting the set of physical computers under consideration. As an alternative, to formalize the notion that a computer's inputs and outputs be "intelligible", here we consider universes having another computer which can consider the first one. We then require that that second computer be able to directly pose binary questions about whether the first computer's prediction correctly corresponds to reality, without relying on any intervening "translational" computer to interpret that first computer. (Note that nothing is being said about whether such a question can be correctly *answered* by the second computer, simply whether it can be *posed* to that computer.) So we wish to be able to ask if that output is one particular value, whether it is another particular value, whether it is one of a certain set of values, etc. Intuitively, this means that the set Q for the second computer must contain binary functions of OUT(.) of the first computer. Finally, we also require that the second computer be similarly intelligible to the first one.

These two requirements are how we impose the intuitive requirement that both computers be "readily intelligible" as predictions concerning reality; they must be readily intelligible and checkable *to each other*. They are formalized and generalized as follows:

**Definition 3:** Consider a physical computer $C \equiv (Q, IN(.), OUT(.))$ and a $\hat{U}$-partition $\pi$. A function from $\hat{U}$ into **B**, f, is an *intelligibility function* (for $\pi$) if

$$\forall \ \hat{u}, \ \hat{u}' \in \ \hat{U}, \pi(\ \hat{u}\ ) = \pi(\ \hat{u}'\ ) \Rightarrow f(\ \hat{u}\ ) = f(\ \hat{u}'\ ).$$

A set F of such intelligibility functions is an *intelligibility set* for $\pi$.

We view any intelligibility function as a question by defining A(f) to be the image of $\hat{U}$ under f. If F is an intelligibility set for $\pi$ and $F \subseteq Q$, we say that $\pi$ is *intelligible* to C with respect to F. If the intelligibility set is not specified, it is implicitly understood to be the set of all intelligibility functions for $\pi$.

We say that two physical computers $C^1$ and $C^2$ are *mutually intelligible* (with respect to the pair $\{F^i\}$) iff both $OUT^2$ is intelligible to $C^1$ with respect to $F^2$ and $OUT^1$ is intelligible to $C^2$ with respect to $F^1$.

Plugging in, $\pi$ is intelligible to C iff $\forall$ intelligibility functions f, $\exists$ q $\in$ $OUT_q$ such that q = f, i.e., such that A(q) = the image of $\hat{U}$ under f, and such that $\forall$ $\hat{u}$ $\in$ $\hat{U}$, q($\hat{u}$) = f($\hat{u}$). Note that since $\pi$ contains at least two elements, if $\pi$ is intelligible to C, $\exists$ $OUT_q$ $\in$ $\{OUT_q\}$ such that A($OUT_q$) = **B**, an $OUT_q$ such that A($OUT_q$) = {0}, and one such that A($OUT_q$) = {1}. Usually we are interested in the case where $\pi$ is an output partition of a physical computer, as in mutual intelligibility.

Intuitively, an intelligibility function for a partition $\pi$ is a mapping from the elements of $\pi$ into **B**. $\pi$ is intelligible to C if Q contains all binary-valued functions of $\pi$, i.e., if C can have posed any question concerning the universe as measured on $\pi$. This flexibility in C ensures that C's output partition isn't "rigged ahead of time" in favor of some particular question concerning $\pi$. Formally, by the surjectivity of OUT(.), demanding intelligibility implies that $\exists$ $\hat{u}'$ $\in$ $\hat{U}$ such that $\forall$ $\hat{u}$ $\in$ $\hat{U}$, $[OUT_q(\hat{u}')](\hat{u})$ = f($\hat{u}$).

In conventional computation IN(.) specifies the question q $\in$ Q we want to pose to the computer (see the example above). In such scenarios, mutual intelligibility restricts how much computation can be "hidden" in $OUT^2$(.) and $IN^1$(.) ($OUT^1$(.) and $IN^2$(.), respectively) by coupling them, so that subsets of the range of $OUT^2$(.) are, directly, elements in the range of $IN^1$(.), without any intervening computational processing.

More prosaically, to motivate intelligibility we can simply note that we wish to be able to pose to $C^1$ any prediction question we can formulate. In particular, this means we wish to be able to pose to $C^1$ any questions concerning well-defined aspects of the future state of $C^2$. Now consider having $C^2$ be a conventional computer based on an underlying physical system. Then we want to be able to predict $C^2$'s output at time $\tau$ as $OUT^2(u_\tau)$. Therefore in addition to any other questions we might want to be able to pose to it, we want to be able to pose to $C^1$ questions involving the value $OUT^2(u_\tau)$ (e.g., is that value $x_1$? $x_1$ or $x_2$? $x_1$ or $x_3$? etc.). This is equivalent to requiring intelligibility.

### iii) Predictable computation

We can now formalize the concept of a physical computer's "making a correct prediction" concerning another computer's future state. We do this as follows:

**Definition 4:** Consider a physical computer C, partition $\pi$, and intelligibility set for $\pi$, F. We say that $\pi$ is *weakly predictable* to C with respect to F iff:

  i) $\pi$ is intelligible to C with respect to F, i.e., $F \subseteq OUT_q$ ;

  ii) $\forall\, f \in F$, $\exists\ IN \in \{IN\}$ that *weakly induces* f, i.e., an IN such that:

  $IN(\hat{u}) = IN$

  $\Rightarrow$

  $OUT_p(\hat{u}) = (A(OUT_q(\hat{u})), OUT_\alpha(\hat{u})) = (A(f), f(\hat{u}))$.

Intuitively, condition (ii) means that for all questions q in F, there is an input state such that if C is initialized to that input state, C's answer to that question q (as evaluated at $\tau$) must be correct. We will say a computer C' with output OUT'(.) is weakly predictable to another if the partition $OUT'_p(.)$ is. If we just say "predictable" it will be assumed that we mean weak predictability.

This definition of predictable is extremely weak. It only concerns those $q \in Q$ that are in F. Also, it doesn't even require that there be a sense in which the information input to C is interpretable as a description of the external universe. (This freedom is what allows us to avoid formalizing the concept of whether some input does or does not "correctly describe" the external universe.) Furthermore, even if the input is interpretable that way, we don't require that it be correct. As an example of this, as in conventional computation {IN} could consist of specifications of t = 0 states of some system S whose future we want to predict, i.e., IN(.) maps the t = 0 state of a physical system underlying our computer to the space of possible t = 0 states of S. But nothing in the definition of 'predictable' requires that $IN(\hat{u})$ *correctly* specifies the values of those initial conditions of S; all that matters is that the resulting prediction be correct. Indeed, we don't even require that $OUT_q(\hat{u}) = q$. Even if the computer gets confused about what question it's answering, we give it credit if it comes up with the correct answer to our question. In addition, we do not even require that the value IN uniquely fixes $OUT_\alpha(\hat{u})$. There may be two $\hat{u}$'s both consistent with IN that nonetheless have different $OUT_\alpha(\hat{u})$ (and therefore correspond to different values of q).

Finally, note that none of the times 0, $\tau$ or T occur in the definition of 'predictable' or in any of the terms going into that definition. Although we motivated the definition as a way to analyze prediction of the future, it actually encompasses a much broader range of kinds of computation. So although our results below do govern prediction of the future, they have many other ramification as well. Most generally, they govern issues concerning sets of potential properties of the universe's history. In particular, they govern whether one can always restrict that history together with the state of a "computer", itself specified in that history, so that that computer correctly guesses which of the properties actually holds. (In this, although it is pedagogically helpful, use of the term "prediction" is a bit misleading.)

Even when there is temporal ordering of inputs, outputs, and the prediction involved in the computation, they need not have T > $\tau$ > 0. We could just as easily have T < $\tau$ < 0 or even T < 0 < $\tau$. So the results presented below will establish the uncomputability *of the past* as well as of the future. They also can be viewed as establishing the fallibility of any observation apparatus and of any control apparatus. These points will be returned to below.

As a formal matter, note that in the definition of predictable, even though f(.) is surjective onto A(f) (cf. Def. 3), it may be that for some IN, the set of values f($\hat{u}$) takes on when $\hat{u}$ is restricted so that IN($\hat{u}$) = IN do not cover all of A(f). The reader should also bear in mind that by surjectivity, $\forall$ IN $\in$ {IN}, $\exists$ $\hat{u}$ $\in$ $\hat{U}$ such that IN($\hat{u}$) = IN.

**iv) Distinguishable computers**

There is one final definition that we need before we can establish our unpredictability results. In our analysis below we will need to have a formal definition of what we mean by having two separate physical computers. The basic idea behind this definition is that just as we require that all input values IN $\in$ {IN} are physically realizable states of a single physical computer, so all pairs of the two computer's inputs values must be physically realizable states of the two physical computers. Intuitively, the computers are not so intertwined that how we can initialize one of them is determined by how we initialize the other. We formalize this as follows:

**Definition 5:** Consider a set of n physical computers $\{C^i \equiv (Q^i, IN^i(.), OUT^i(.)) : i = 1, ..., n\}$. We say $\{C^i\}$ is (*input*) *distinguishable* iff $\forall$ n-tuples $(IN^1 \in \{IN^1\}, ..., IN^n \in \{IN^n\})$, $\exists \, \hat{u} \in \hat{U}$ such that $\forall$ i, $IN^i(\hat{u}) = IN^i$ simultaneously.

We say that $\{C^i\}$ is *pairwise* (*input*) *distinguishable* if any pair of computers from $\{C^i\}$ is distinguishable, and will sometimes say that any two such computers $C^1$ and $C^2$ "are distinguishable from each other". We will also say that $\{C^i\}$ is a *maximal* (pairwise) distinguishable set if there are no physical computers $C \notin \{C^i\}$ such that $C \cup \{C^i\}$ is a (pairwise) distinguishable set.

## 2. THE UNCOMPUTABILITY OF THE FUTURE

### i) The impossibility of posing arbitrary questions to a computer

Our first result does not even concern the accuracy of prediction. It simply states that for any pair of physical computers there are *always* binary-valued questions about the state of the universe that cannot even be posed to at least one of those physical computers. In particular, this is true if the second computer is a copy of the first one, or even if it is the same as the first one. (The result does not rely on input-distinguishability of the two computers — a property that obviously does not describe the relationship between a computer and itself.) This impossibility holds no matter what the cardinality of the set of questions that can be posed to the computers (i.e., no matter what the cardinality of $\{IN\}$ and/or Q). It is also true no matter how powerful the computers (and in particular holds even if the computers are more powerful than a Turing Machine), whether the computers are analog or digital, whether the universe is classical or quantum-mechanical, whether or not the computers are quantum computers, and even whether the computers are subject to physical constraints like the speed of light. In addition the result does not rely on chaotic dynamics in any manner. All that is required is that the universe contain two (perhaps identical, perhaps wildly different) physical computers.

**Theorem 1:** Consider any pair of physical computers $\{C^i : i = 1, 2\}$. Either $\exists$ finite intelligibility set $F^2$ for $C^2$ such that $C^2$ is not intelligible to $C^1$ with respect to $F^2$, and/or $\exists$ finite intelligibility set $F^1$ for $C^1$ such that $C^1$ is not intelligible to $C^2$ with respect to $F^1$.

**Proof:** Hypothesize that the theorem is false. Then $C^1$ and $C^2$ are mutually intelligible $\forall$ finite $F^1$ and $F^2$. Now the set of all finite $F^2$ includes any and all intelligibility functions for $C^2$, i.e., any and all functions taking $\hat{u}$ to a bit whose value is set by the value $OUT^2(\hat{u})$. The set of those functions can be bijectively mapped to the power set $2^{\{OUT^2\}}$. So $F^2 \subseteq Q^1 \Rightarrow o(Q^1) \geq o(2^{\{OUT^2\}})$. However $o(\{OUT^2\}) \geq o(Q^2)$, since $\{OUT^2\}$ contains all possible specifications of a $q^2 \in Q^2$. Therefore $o(Q^1) \geq o(2^{Q^2})$. But it is always true that $o(2^A) > o(A)$ for any set A, which means in particular that $o(2^{Q^2}) > o(Q^2)$. Accordingly, $o(Q^1) > o(Q^2)$. Similarly though, $o(Q^2) > o(Q^1)$. Therefore $o(Q^1) > o(Q^1)$, which is impossible. **QED**.

Ultimately, Thm. 1 holds due to our requiring that our physical computer be capable of answering more than one kind of question about the future state of the universe. To satisfy this requirement q cannot be pre-fixed. (In conventional computation, it is specified in the computer's input.) But precisely because q is not fixed, for the computer's output of $\alpha$ to be meaningful it must be accompanied by specification of q; the computer's output must be a well-defined region in $\hat{U}$. It is this need to specify q as well as $\alpha$, ultimately, which means that one cannot have two physical computers both capable of being asked arbitrary questions concerning the output of the other.

Thm. 1 reflects the fact that while we do not want to have C's output partition "rigged ahead of time" in favor of some single question, we also cannot require too much flexibility of our computer. It is necessary to balance these two considerations. Before analyzing prediction of the future, to circumvent Thm. 1 we must define a restricted kind of intelligibility set to which Thm. 1 does not apply. This is a set of functions whose value does not depend on the question component of OUT, only on the answer component. Intuitively, restricting ourselves to these kinds of intelligibility sets means we are only requiring that the predicted partition *label* of one physical computer be directly readable on the other computer's input, not that the full partition of the first computer's question also be directly readable.

Recall that for any f that is an intelligibility function for (the output partition of) some computer C, $\forall$ $\hat{u}$, $\hat{u}' \in \hat{U}$, OUT( $\hat{u}$ ) = OUT( $\hat{u}'$ ) implies that f( $\hat{u}$ ) = f( $\hat{u}'$ ). So for such an f, the joint condition [OUT$_q$( $\hat{u}$ ) = OUT$_q$( $\hat{u}'$ )] $\wedge$ [OUT$_\alpha$( $\hat{u}$ ) = OUT$_\alpha$( $\hat{u}'$ )] implies that f( $\hat{u}$ ) = f( $\hat{u}'$ ). We consider f's that obey weaker conditions:

**Definition 6:** An intelligibility function f for an output partition OUT(.) is *question-independent* iff $\forall$ $\hat{u}$, $\hat{u}' \in \hat{U}$:

$$OUT_p( \hat{u} ) = OUT_p( \hat{u}' )$$

$$\Rightarrow$$

$$f( \hat{u} ) = f( \hat{u}' ).$$

An intelligibility set as a whole is question-independent if all its elements are.

We write $C^1 > C^2$ (or equivalently $C^2 < C^1$) and say simply that $C^2$ is (weakly) *predictable* to $C^1$ (or equivalently that $C^1$ *can predict* $C^2$) if $C^2$ is weakly predictable to $C^1$ for all question-independent finite intelligibility sets for $C^2$.

Similarly, from now on we will say that $C^2$ is *intelligible* to $C^1$ without specification of an intelligibility set if $C^2$ is intelligible to $C^1$ with respect to all question-independent finite intelligibility sets for $C^2$.

Intuitively, f is question-independent if its value does not vary with q among any set of q all of which share the same A(q). As an example, say our physical computer is a conventional digital workstation. Have a certain section of the workstation's RAM be designated the "output section" of that workstation. That output section is further divided into a "question subsection" designating (i.e., "containing") a q, and an "answer subsection" designating an $\alpha$. Say that for all q that can be designated by the question subsection A(q) is a single bit, i.e., we are only interested in binary-valued questions. Then for a question-independent f, the value of f can only depend on whether the answer subsection contains a 0 or a 1. It cannot vary with the contents of the question subsection.

As a formal example of question-independent intelligibility, say our computer has questions q for which A(q) = **B**, questions q for which A(q) = {0}, and q for which A(q) = {1}, but no others. Then there are four distinct subsets of $\hat{U}$, which mutually cover $\hat{U}$, defined by the four equations $\text{OUT}_p(\hat{u}) = (\mathbf{B}, 1)$, $\text{OUT}_p(\hat{u}) = (\mathbf{B}, 0)$, $\text{OUT}_p(\hat{u}) = (\{1\}, 1)$, and $\text{OUT}_p(\hat{u}) = (\{0\}, 0)$. (The full partition OUT(.) is a refinement of this 4-way partition, whereas this 4-way partition need not have no relation with the partitions making up each q in Q.) So a question-independent intelligibility function of our computer is any **B**-valued function of which of these four subsets a particular $\hat{u}$ falls into.

In terms of the first of the motivations we introduced for requiring intelligibility, requiring question-independent intelligibility means we only require each computer's *answer* to be readily intelligible to the other one. We are willing to forego having the question that each computer thinks it's answering also be readily intelligible to the other one. Alternatively, we can define a "partial computer" as a modified kind of computer whose variable OUT is only A(q) and $\alpha$ (rather than q and $\alpha$). Intelligibility in the sense originally defined, applied to a partial computer, is exactly equivalent to applying question-independent intelligibility to a full computer. In particular, the set of all question-independent intelligibility functions of any output partition OUT(.) equals the set of all intelligibility functions of the partial computer output partition $\text{OUT}_p(.)$.

Thm. 1 does not hold if we restrict attention to question-independent intelligibility sets. As an example, both of our computers could have their output answer subsections be a single bit, and both could have their Q contain all four Boolean questions about the state of the other computer's output answer bit. (Those are the following functions from $\hat{u} \in \hat{U} \rightarrow \mathbf{B}$: Is u such that the other computer's output bit is 1? 0? 1 and/or 0? Neither 1 nor 0?) So the Q of both computers contains all possible question-independent intelligibility sets for the other computer.

The following example establishes that there are pairs of input-distinguishable physical computers $\{C^1, C^2\}$ in which $C^2$ is predictable to $C^1$:

**Example 2:** $Q^2$ consists of a single question, one which is a binary partition of $\hat{U}$ so that $A(OUT^2_q(\hat{u})) = \mathbf{B}$ always. Since $OUT^2(.)$ is surjective, the image of $\hat{U}$ under $OUT^2_\alpha(.)$ is all of $\mathbf{B}$. $Q^1$ has four elements given by the four logical functions of the bit $OUT^2_\alpha(\hat{u})$. (Note these are the four intelligibility functions for $C^2$.) Have $IN^1(.) = OUT^1_q(.)$, so that $\{IN^1\}$ contains four elements corresponding to those four possible questions concerning $OUT^2_\alpha$. Next, have $OUT^1_\alpha(\hat{u})$ $= [OUT^1_q(\hat{u})](\hat{u}) \forall \hat{u} \in \hat{U}$. Then for any of the four intelligibility functions for $C^2$, q, $\exists IN^1 \in \{IN^1\}$ such that $IN^1(\hat{u}) = IN^1 \Rightarrow [A(OUT^1_q(\hat{u})) = A(q)]$ ∧ $[OUT^1_\alpha(\hat{u}) = q(\hat{u})]$; simply choose $IN^1 = q$, so that $IN^1(\hat{u}) = IN^1 \Rightarrow OUT^1_q(\hat{u}) = q$. Finally, to ensure distinguishability, if there are multiple $IN^2$ values, have each one occur for at least one $\hat{u}$ in each of the subregions of $\hat{U}$ given by the partition $IN^1(.)$.

To ensure surjectivity of $OUT^1(.)$, we could have $IN^1(.)$ subdivide each of the two sets (one set for each value of $OUT^2_\alpha$) $\{ \hat{u} \in \hat{U} : OUT^2_\alpha(\hat{u}) = OUT^2_\alpha \}$ into four non-empty subregions, one for each $IN^1$ value. So $(IN^1(\hat{u}), OUT^2_\alpha(\hat{u}))$ are two-dimensional coordinates of a set of disjoint regions that form a rectangular array covering $\hat{U}$. This means that $\hat{u} \rightarrow (IN^1(\hat{u}), OUT^2_\alpha(\hat{u}))$ is surjective onto $\{IN^1\} \times \{OUT^2_\alpha\}$, so that for any $OUT^1_\alpha$ and intelligibility function for $C^2$, q, there is always a value of $IN^1$ that both induces the correct prediction for that function q and is consistent with that $OUT^2_\alpha$.

The following variant of Ex. 2 establishes that we could have yet another computer $C^3$ that predicts $C^2$ but that is also distinguishable from $C^1$:

**Example 2':** Have $Q^3 = Q^1$, $\{IN^3\} = \{IN^1\}$, $OUT^3{}_q(.) = IN^3(.)$, $OUT^3{}_\alpha(\hat{u}) = [OUT^3{}_q(\hat{u})](\hat{u})$ $\forall \hat{u} \in \hat{U}$, and have $IN^3(.)$ subdivide $IN^1(.)$ so that all four values of $IN^3$ can occur with each value of $IN^1$. In general, as we vary over all $\hat{u} \in \hat{U}$ and therefore over all $(IN^1, IN^3)$ pairs, the pair of intelligibility function that $C^1$ is predicting will separately vary from those that $C^3$ is predicting, in such a way that all $2^4$ pairs of intelligibility functions for $C^2$ are answered correctly for some $\hat{u}$ $\in \hat{U}$.

In addition, we can have a computer $C^4$, distinguishable from both $C^1$ and $C^2$, where $C^4 > C^1$, so that $C^4 > C^1 > C^2$. We can do this either with $C^4 > C^2$ or not, as the following variant of Ex. 2 demonstrates:

**Example 2'':** Have $OUT^4{}_q(.) = IN^4(.)$, $OUT^4{}_\alpha(\hat{u}) = [OUT^4{}_q(\hat{u})](\hat{u})$ $\forall \hat{u} \in \hat{U}$, and $\{IN^4\} = \{OUT^4{}_q\}$ equals the set of all $2^4$ question-independent intelligibility functions for $C^1$. (There are four possible $OUT^1{}_p$: $\{(\{0\}, 0), (\{1\}, 1), (\mathbf{B}, 0), (\mathbf{B}, 1)\}$.) Ensure surjectivity of $OUT^4(.)$ by having each region of constant $OUT^4{}_q(\hat{u})$ overlap each region of constant $OUT^1{}_p(\hat{u})$. This establishes that $C^4 > C^1$. Distinguishability would then hold if $IN^4(.)$ subdivides $IN^1(.)$ so that all 16 values of $IN^4$ can occur with each value of $IN^1$.

In this setup, $C^2$ may or may not be predictable to $C^4$. To see how it may not be, consider the case where $\{IN^2\}$ is a single element (so distinguishability with $C^2$ is never an issue). Have $IN^4(.)$ be a refinement of $OUT^2{}_\alpha(.)$, in that each $IN^4$ value can only occur with one or the other of the two $OUT^2{}_\alpha$ values. So each $IN^4$ value delineates a "horizontal strip" of constant $OUT^2{}_\alpha(\hat{u})$, running across all four values of $IN^1(\hat{u})$. (Since $IN^1(\hat{u}) = OUT^1{}_q(\hat{u})$, and $OUT^1{}_\alpha(\hat{u}) = (OUT^1{}_q(\hat{u}))(\hat{u})$, $OUT^1{}_\alpha(\hat{u}) = (IN1(\hat{u}))(\hat{u})$, so specifying the value of $IN^1(\hat{u})$ specifies $OUT^1{}_p(\hat{u})$, and each strip crosses all four $OUT^1{}_p$ values, as was stipulated above.)

Now choose the strip with $A(OUT^4{}_q(\hat{u})) = A(IN^4(\hat{u})) = \{0\}$ to have coordinate $OUT^2{}_\alpha(\hat{u})$ $= 1$, and the strip with $A(OUT^4{}_q(\hat{u})) = \{1\}$ to have coordinate $OUT^2{}_\alpha(\hat{u}) = 0$. In the remaining fourteen strips, $OUT^4_\alpha(\hat{u})$ is not constant, and therefore is not a single-valued intelligibility function of the associated (constant) value of $OUT^2_p(\hat{u})$. In both of those two strips though, $OUT^4_\alpha(\hat{u})$ is the opposite of $OUT^2_\alpha(\hat{u})$. So no $IN^4$ value induces the identity question-independent intelligibility function for $C^2$: $\hat{u} \rightarrow OUT^2_\alpha(\hat{u})$, i.e., no $IN^4$ induces $OUT^4_p(\hat{u}) = (\mathbf{B}, OUT^2_\alpha(\hat{u}))$. Accordingly, $C^4$ does not predict $C^2$.

In other instances though, both $C^2$ and $C^1$ are predictable to $C^4$. To have this we need only subdivide $\{IN^4\}$ and $\{OUT^4\}$ into two portions, $(\{IN^4\}_A, \{OUT^4\}_A)$, and $(\{IN^4\}_B, \{OUT^4\}_B)$, which divide $\hat{U}$ in two. The first of these portions is used for predictions concerning $C^2$, as in Ex. 2; each region of constant $IN^4(\hat{u})$ is a subset of a region of constant $IN^1(\hat{u})$ overlapping both $OUT^2_\alpha(\hat{u})$. The second is used for predictions concerning $C^1$, as just above. It consists of horizontal strips extending over that part of $\hat{U}$ not taken up by the regions with $IN^4(\hat{u}) \in \{IN^4\}_A$. So $\{IN^4\}_A = \{OUT^4_q\}_A$ contains four elements, and $\{IN^4\}_B = \{OUT^4_q\}_B$ contains sixteen, which means that $\{IN\} = \{OUT\}$ contains twenty elements, all told. Distinguishability is ensured by having $IN^4$ take on all its possible values within any subset of $\hat{U}$ over which both $IN^1(.)$ and $IN^2(.)$ are constant.

## ii) The impossibility of assuredly correct prediction

Even if we can pose all the questions in some set to a computer, that says nothing about whether by appropriate choice of input the computer can always be assured of correctly answering any question from that set. In fact, even if we restrict attention to question-independent intelligibility sets, no physical computer can be assuredly correct in its predictions concerning the future.

Whereas the impossibility expressed by Thm. 1 follows from cardinality arguments and the power set nature of intelligibility sets, the impossibility of assuredly correct prediction follows from the presence of the negation operator in a (question-independent) intelligibility set. As an example of the logic underlying the proof, consider a pair of computers predicting the future, both of whose output answer subsections are binary. Have one of the two computers predict the other's output bit, whereas that other computer predicts the negation of the first one's output bit. Since both computers' output calculations must halt by $\tau$, they will contradict each other when the prediction time arrives. Therefore they cannot both be correct in their predictions.

To formalize this, first note that for any partition $\pi$ containing at least two elements, there exists an intelligibility function f for $\pi$ with A(f) = **B**, an intelligibility function f with A(f) = {1}, and an intelligibility function f with A(f) = {0}. By exploiting the surjectivity of output partitions, we can extend this result to concern such partitions. This is formally established in the following lemma, which holds whether or not we assume partitions are binary:

**Lemma 1:** Consider a physical computer $C^1$. If $\exists$ any output partition $OUT^2$ that is intelligible to $C^1$, then $\exists$ $q^1 \in Q^1$ such that A($q^1$) = **B**, a $q^1 \in Q^1$ such that A($q^1$) = {0}, and a $q^1 \in Q^1$ such that A($q^1$) = {1}.

**Proof:** Since {$OUT^2$} is non-empty, {$OUT^2_q$} is non-empty. Pick some $q^* \in$ {$OUT^2_q$} having at least two elements. (By definition of physical computer, there is at least one such $q^*$.) Construct any binary-valued function $f^{*2}$ of $\alpha \in$ A($q^*$) such that there exists at least one $\alpha$ for which $f^{*2}(\alpha)$ = 0 and at least one for which $f^{*2}(\alpha)$ = 1. Define an associated function $f^{*2}(\hat{u})$ = $f^{*2}(OUT^2_\alpha(\hat{u}))$ if A($OUT^2_q(\hat{u})$) = A($q^*$), 0 otherwise. By the surjectivity of $OUT^2$(.), $\forall \alpha \in$ A($q^*$), $\exists \hat{u}$ such that both $OUT^2_q(\hat{u})$ = $q^*$ and $OUT^2_\alpha(\hat{u})$ = $\alpha$. Therefore $\exists \hat{u}$ such that $f^{*2}(\hat{u})$ = 1, and $\exists \hat{u}$ such that $f^{*2}(\hat{u})$ = 0.

This establishes, by construction, that there is a question-independent intelligibility function for $C^2$ that takes on both the value 1 and the value 0, $f^{*2}$. So by our hypothesis that $C^2$ is intelligible to $C^1$ with respect to any question-independent intelligibility function for $C^2$, we know that $f^{*2} \in Q^1$. Moreover, viewed as a question, A($f^{*2}$) = **B**. So, we have established that $Q^1$ contains a binary valued function.

Next, note that the function $\hat{u} \in \hat{U} \rightarrow 1$ is always a question-independent intelligibility function for $C^2$, as is the function $\hat{u} \in \hat{U} \rightarrow 0$. Again using surjectivity, we see that A for these two functions are {1} and {0}, respectively. **QED.**

We can now establish our central theorem:

**Theorem 2**: Consider any pair of distinguishable physical computers $\{C^i : i = 1, 2\}$. It is not possible that both $C^1 > C^2$ and $C^1 < C^2$.

**Proof:** Hypothesize that the theorem is false. Then $C^1$ and $C^2$ are mutually predictable for all pairs of question-independent intelligibility functions (one function for each computer), and therefore mutually intelligible for them as well. Therefore Lemma 1 applies. Using the surjectivity of $OUT^2(.)$, this means that $\exists q^2 \in Q^2$ such that $A(q^2) = \mathbf{B}$ and such that there both exists a $\hat{u} \in \hat{U}$ such that $OUT^2(\hat{u}) = (q^2, 0)$ and a $\hat{u} \in \hat{U}$ such that $OUT^2(\hat{u}) = (q^2, 1)$. By similar reasoning, $\exists q^1 \in Q^1$ such that $A(q^1) = \mathbf{B}$, and such that there both exists a $\hat{u} \in \hat{U}$ such that $OUT^1(\hat{u}) = (q^1, 0)$ and a $\hat{u} \in \hat{U}$ such that $OUT^1(\hat{u}) = (q^1, 1)$.

Consider the function of $\hat{u} \in \hat{U}$ whose value is 1 if $OUT^2_p(\hat{u}) = (\mathbf{B}, 1)$, 0 otherwise. Like $q^1$, this is a question-independent intelligibility function for $C^2$, and by our argument just above, we know it is surjective onto $\mathbf{B}$. Again using mutual intelligibility, this intelligibility function is a $q \in Q^1$, $q^{*1}$. Intuitively, this q for $C^1$ is just the bit of $C^2$'s output answer, for those cases where that answer is binary. (Our proving surjectivity establishes that there actually are such cases where the space of answers is binary, and furthermore that among such cases both output answers arise.) Similarly, $\exists q \in Q^2$, $q^{*2}$, such that $q^{*2}(\hat{u}) = 1$ if $OUT^1_p(\hat{u}) = (\mathbf{B}, 0)$, 1 otherwise. Intuitively, this q for $C^2$ is just the negation of the bit of $C^1$'s output answer.

By hypothesis our computers are mutually predictable with respect to any two finite intelligibility sets. Therefore they are mutually predictable with respect to the two (single-element) intelligibility sets for $C^2$ and $C^1$, $q^{*1}$, and $q^{*2}$, respectively. Therefore $\exists IN^2$ such that $IN^2(\hat{u}) = IN^2$ implies that $OUT^2_p(\hat{u}) = (\mathbf{B}, q^{*2}(\hat{u}))$. Similarly, $\exists IN^1$ such that $IN^1(\hat{u}) = IN^1$ implies that $OUT^1_p(\hat{u}) = (\mathbf{B}, q^{*1}(\hat{u}))$. But since our computers are input-distinguishable, $\exists \hat{u}$ for which both $IN^2(\hat{u}) = IN^2$ and $IN^1(\hat{u}) = IN^1$. Therefore $\exists \hat{u}$ for which $A(OUT^2_q(\hat{u})) = A(OUT^1_q(\hat{u})) = \mathbf{B}$, $OUT^2_\alpha(\hat{u}) = q^{*2}(\hat{u})$, and $OUT^1_\alpha(\hat{u}) = q^{*1}(\hat{u})$.

Plugging in, for that $\hat{u}$, $A(OUT^2_q(\hat{u})) = \mathbf{B}$, and $OUT^2_\alpha(\hat{u}) = 1$ if $OUT^1_p(\hat{u}) = (\mathbf{B}, 0)$, 1 otherwise. Similarly, $A(OUT^1_q(\hat{u})) = \mathbf{B}$, and $OUT^1_\alpha(\hat{u}) = 1$ if $OUT^2_p(\hat{u}) = (\mathbf{B}, 1)$, 0 otherwise. Plugging in again, we have $OUT^2_\alpha(\hat{u}) = 1$ if $OUT^2_\alpha(\hat{u}) \neq 1$, 0 otherwise. This contradiction establishes that our hypothesis is wrong, which establishes the theorem. **QED.**

Restating it, Thm. 2 says that either $\exists$ finite question-independent intelligibility set for $C^1$, $F^1$, such that $C^1$ is not predictable to $C^2$ with respect to $F^1$, and/or $\exists$ finite question-independent intelligibility set for $C^2$, $F^2$, such that $C^2$ is not predictable to $C^1$ with respect to $F^2$.

Thm. 2 holds no matter how large and powerful our computers are; it even holds if the "physical system underlying" one or both of our computers is the whole universe. It also holds if instead $C^2$ is the rest of the physical universe external to $C^1$. As a particular instance of this latter case, the theorem holds even if $C^1$ and $C^2$ are physically isolated from each other $\forall$ t > 0. (Results similar to Thm. 2 that rely on physical coupling between the computers are presented in [30].)

Rather than viewing it as imposing limits on computers, Thm. 2 can instead be viewed as imposing limits on the computational capabilities of the universe as a whole. In this perspective that theorem establishes that the universe cannot support parallel computation in which all the nodes are sufficiently powerful to correctly predict each other's behavior.

### iii) Implications of the impossibility of assuredly correct prediction for a single computer

Let C be a computer supposedly capable of correctly predicting the future of any system S if information concerning the initial state of S is provided to C, as in example 1 above. Assume that C is not so powerful that the universe is incapable of supporting a copy of C in addition to the original. (This is certainly true of any C conceivably built by humans.) Have S be such a copy of C. We assume that for any pair of t = 0 input values, there is at least one world-line of the universe in which C's input is one of those values and the other value constitutes the input of C's copy (i.e., we have input-distinguishability).

Applying Thm. 1 to our two computers, we see that there is a finite intelligibility set that is not intelligible to C, i.e., there are questions concerning an S that cannot even be posed to C. (More formally, there is either such a set for C or for its copy, S.) In addition, by Thm. 2, there is a finite question-independent (and therefore potentially pose-able) intelligibility set concerning S that is not predictable to C. In other words, there must be a question-independent intelligibility function concerning S that C predicts incorrectly, no matter what the input to C (assuming the function can even be posed to C at all).

The binary partition over $U_T$ induced by this unpredictable intelligibility function constitutes a question concerning the time T state of S. In addition every one of the set of potential inputs to C corresponds to a subset of $U_0$, and therefore corresponds to a subset of the possible states of C's "input section" at time 0. (In Ex. 1, IN(.) is set up so that every element in {IN} corresponds to one and only one state of C's input section at time 0.) Similarly, every output of C corresponds to a subset of $U_\tau$ and therefore a subset of the possible states of C's "output section" at time $\tau$. Accordingly, our result means that there is no input to C at time 0 that will result in C's output at time $\tau$ having the correct answer to our question concerning the time T state of S. For $0 < \tau < T$, this constitutes a formal proof that no computer can predict the future faster than it occurs. (Or more precisely, that the universe cannot support more than one such computer.) As mentioned previously, the result also holds for $T < 0$ however, in which case it denies the possibility of assuredly correct prediction of the past.

While these results hold if C and S are isolated from one another $\forall$ t > 0, they also hold if C and S are coupled at such times. Indeed, they hold no matter what the form of such coupling. So in particular, we can have S be a copy of C that is coupled with the original by having that original "observe" the copy's output section. Doing this, our result establishes the impossibility of a device C that can take specification of any characteristic of the universe as input, observe the value of that characteristic, then report that value and have the value still be true at the time of reporting. (Note that when a computer is used for observation, its input will in general not uniquely fix its output, unlike the case with prediction discussed in Ex. 1.) This impossibility holds independent of considerations of light-cones and the like, and in fact holds just as well in a universe with c = ∞ as it does in ours. (Alternatively, the time at which the characteristic is to be observed can be specified in the computer's input, and therefore can be far enough into the future so that C's light-cone can intersect with that of the characteristic.) In all this, Thm. 2 establishes the impossibility of a general-purpose observation apparatus.

Moreover, there is nothing in the math that forces C to play a "passive observational role" in the coupling with S. So we can just as well view Thm. 2 as establishing the impossibility of an apparatus capable of ensuring that there is no discrepancy between the time T value in some physical computer C's output section and an associated characteristic of a system S external to C. There is no such thing as an infallible general-purpose controller.

Whether used for prediction, observation, or control, one can "start" our computer at any time before t = 0 (i.e., give the computer a potentially semi-infinite "running start") and our impossibility results still hold. In addition, it is worth noting that in the context of prediction or observation, these impossibility results hold even if one tries to have the input to the computer explicitly contains the correct value of the prediction or observation. (Since the universe is single-valued / deterministic, such a value must exist.) Impossibility also obtains if the input is stochastic (since it holds for each input value individually). Similarly, although we are primarily interested in computers that run programs which were specifically designed to try to achieve our computational task, nothing in the theorem requires this. *Whatever the program*, our result shows that the computer it runs on must have output which never equals the correct answer.

Thm. 2 applies even if we consider human beings, perhaps individually or in a group, perhaps using physical computational aids, perhaps building special-purpose physical devices, to be the "underlying physical computers". Even in a classical, non-chaotic, finite universe, there cannot be two scientists both of whom are infallible in their calculations (or even their observations) concerning the universe. This is even true if we accept Penrose's thesis that somehow quantum gravity imbues human beings with extraordinary computational powers. No restrictions are set in the theorem on how the computer operates, and there are no explicit assumptions about the computational power of either the computer or of the universe. Indeed, even if the computer is infinitely fast and/or dense, or powerful enough to solve the Halting Problem, the theorem still holds.

Impossibility results that are in some senses even stronger than those associated with Thm. 2 hold when we do not restrict ourselves to distinguishable computers, as we do in Thm. 2. For example, some of those results establish the impossibility of a computer C's assuredly prediction even if C is so powerful that the universe is not capable of having more than one copy of C. See the discussion in the next paper of prediction complexity.

## FUTURE WORK AND DISCUSSION

Any results concerning physical computation should, at a minimum, apply to the computer lying on a scientist's desk. However that computer is governed by the mathematics of deterministic finite automata, not that of Turing machines. In particular, the impossibility results concerning Turing machines rely on infinite structures that do not exists in any computer on a scientist's desk. Accordingly, it is hard to see why those results should be relevant to a general theory of physical computers.

On the other hand, when one carefully analyzes actual computers that perform calculations concerning the physical world, one uncovers a mathematical structure governing those computers that is replete with its own impossibility results. While much of that structure parallels Turing machine theory, much of it has no direct analogue in that theory. For example, it has no need for structures like tapes, moveable heads, internal states, read/write capabilities, and the like, none of which have any obvious importance to the laws of quantum mechanics and general relativity. Indeed, when the underlying concepts are stripped down to their essentials, one does not even need to identify a "computer" with a particular localized region of space-time, never mind one with heads and the like. In place of all those concepts, one has several partitions over the space of all worldlines of the universe. Those partitions constitutes a computer's inputs, the questions it is addressing, and its outputs. The impossibility results of physical computation concern the relation of those partitions. Computers in the conventional, space-time localized sense (the box on your desk) are simply special examples, with lots of extra restrictions that are unnecessary in the underlying mathematics; the general definition of a "physical computer" has no such restrictions.

One can use this definition of a physical computer to establish many restrictions on what is and is not possible to compute concerning the physical world. The first result is that there cannot be a computer to which one can even pose all possible questions concerning the physical world. In light of this result attention is restricted to a subset of questions, all of whose members can actually be posed to physical computers. Consideration is then focussed on computers that are not so powerful as to be unique, i.e., computers that can exist in multiple renditions in the physical universe. It is shown that no such computer can correctly answer all of the computational questions concerning the physical universe that can be posed to it.

This central result has many implications. The first is that it is impossible for a computer to take a state of any system as an input, and then always correctly predict the future state of that system before it occurs. There must be mistakes made. Loosely speaking, this means that Laplace was wrong: even if the universe were a giant clock, he would not be able to reliably predict its future state before it occurred, no matter how smart he was. Phrased differently, regardless of noise levels and the dimensions and other characteristics of the underlying attractors of the physical dynamics of various, there cannot be a time-series prediction algorithm [8] that is always correct in its prediction of the future state of such systems.

The central result follows solely from the relations among the partitions that jointly specify any physical computer. In particular, it does not rely on physical properties of the universe, like its quantum nature, the finiteness of the speed of light, the chaotic state of its subcomponents, or the like. Nor does it make any assumptions concerning a localized physical that may underlie the physical computer, the Chomsky hierarchy characterization of the computer, or the like. One can exploit this breadth of the definition of "computer" to extend the central result to address many other scenarios besides time-series prediction. In particular, that definition places no restrictions on whether the computer and the system it is predicting are physically coupled. Accordingly, that central result means that one cannot build a general-purpose observation device that always cor-rectly answers an observational question concerning an arbitrary physical system. Similarly, because of the freedom to have the "computer" be coupled to the external system, the central result means that one cannot build a general-purpose infallible control device, i.e., one able to induce a desired state in any specified physical system. These two corollaries of the central result holds independent of concerns about causal relations and light cones; it holds even for an infinite speed of light.

The central result serves as one of the foundations for a mathematics of how sets of computers can be related to one another. For example, say we loosen the assumption that any computer that can exist in our universe can exist in more than rendition. Then the central result can be used to establish that the computability relationship among all computers constitutes a directed acyclic graph. In addition, there is at most one computer that can correctly compute arbitrary questions concerning all others, i.e., that can infallibly predict/observe/control all others. In addition to analyzing the mathematics associated with such "god computers", one can investigate analogues of the Halting theorem of conventional computer science. That investigation results in a natural complexity measure, one that is analogous to algorithmic information complexity. Unlike algorithmic information complexity however, the physical computation analogue is uniquely defined, with no freedom analogous to that in algorithmic information complexity of varying the underlying universal Turing machine. All these issues and many others are discussed in the second of this pair of papers.

Future work related to the central result includes investigating the following issues:

i) How must the definitions and associated results be modified for analog computers (so that one is concerned with amounts of error rather than whether there is an error)? What about if one is calculating the future state of a stochastic system, i.e., if one is predicting a probability distribution?

ii) Are there any modifications to the definitions that would be more appropriate for quantum systems? If so, how are the ensuing results different for quantum systems? (As an example of such a modification, one might want to allow sufficient time between T and $\tau$ to not run into difficulties due to the Heisenberg uncertainty principle.)

iii) How are the results modified if one is concerned with probabilities of erroneous prediction rather than just worst-case analysis of whether there can possibly be erroneous prediction? In particular, how must the results be modified if prediction doesn't involve a bit of whether the universe's actual worldline is or isn't in some particular subset of all such worldlines (a partition element), but instead involves full-blown probability distributions over the full set of all worldlines?

iv) Find the exact point of failure  — which according to (1) and (2) must exist —  of the intuitive argument "If the computer is simply a sufficiently large and fast Hamiltonian evolution approximator, then it can emulate any finite classical non-chaotic system".

v) A related issue is whether any time a computer actually tries to perform a computation of the sort invoked in the proofs of (1) and (2) it is forced to a chaotic trajectory, even though evolution of the combined C-A system in the overall phase space isn't chaotic (i.e., the region of that phase space with positive Lyaponov exponent has measure 0)? Or is it perhaps instead the case that the only kinds of computers one might nominate as capable of predicting any external system are themselves everywhere chaotic? If this were the case, then in both the proof of (1) and of (2), one is trying to predict a chaotic system; might that be the answer to the question in (iv) of why those predictions must be in error? (As an aside, note that if it were the case that only chaotic systems could conceivably function as universal predictors it would rule out the possibility of 100% efficient error-correcting computers as such universal predictors, since they are designed to not be chaotic.)

There are reasons to believe this is not the case. For example, the results in this paper don't simply say that the computer is unreliable, sometimes getting the correct answer, sometimes the wrong answer. Rather those results say that there is a scenario where the computer is *always* wrong. This makes it hard to see how the "explanation" for these results could lie in chaotic or quantum mechanical properties of the computers - one would expect such properties to give unre-liability, not reliable incorrectness.

vi) As mentioned in the introduction, there is a large body of work showing how to embed TM's in physical systems. One topic for future work is following an analogous program in the domain of physical computation, for example by investigating what physical systems support copies of any element of various sets of physical computers.

## FOOTNOTES

[1] To "remember", in the present, an event from the past, formally means "predicting" that event accurately (i.e., retrodicting the event), using only information from the present. Such retrodiction relies crucially on the second law. Hence, the temporal asymmetry of the second law causes the temporal asymmetry of memory (we remember the past, not the future). That asymmetry of memory in turn causes the temporal asymmetry of the psychological arrow of time. "Memory systems theory" refers to the associated physics of retrodiction; it is the thermodynamic analysis of systems for transferring information from the past to the present. See [28].

**ACKNOWLEDGMENTS:** This work was done under the auspices of the Department of Energy, the Santa Fe Institute, and the National Aeronautics and Space Administration. I would like to thank Bill Macready, Cris Moore, Paul Stolorz, Tom Kepler and Carleton Caves for interesting discussion.